\documentclass[10pt,letterpaper,twocolumn]{article} 
\usepackage{ol2}
\usepackage[draft]{hyperref}
\usepackage{amsmath}

\begin{document}

\twocolumn[ 

\title{Interference-induced peak splitting in EUV superfluorescence}


\author{Ni Cui,$^{1,2}$ Christoph H. Keitel, $^{1}$ and Mihai Macovei, $^{1,*}$}

\address{
$^1$Max-Planck-Institut f\"{u}r Kernphysik, Saupfercheckweg 1, D-69117 Heidelberg, Germany
\\
$^2$Department of Physics, College of Science and Engineering, Jinan University, Huangpu Avenue West 601, 510632
Guangzhou, China\\
$^*$Corresponding author: mihai.macovei@mpi-hd.mpg.de\\
}
\begin{abstract}
We investigate the laser-induced quantum interference in EUV superfluorescence occurring 
in a dense gas of $\Lambda$-type helium atoms coupled by a coherent laser field in the visible 
region. Due to the constructive interatomic and intraatomic interferences, the superfluorescence 
can split in two pulses conveniently controlled by the gas density and intensity of the driving field, 
suggesting potential applications for pump-probe experiments.
\end{abstract}

 ] 

\noindent For a collection of $N$ initially excited atoms the constructive interatomic 
interference induces the atoms to radiate cooperatively as a short intense pulse with 
intensity proportional to $N^{2}$ and pulse duration proportional to $N^{-1}$. 
Meanwhile this effect is well known as superfluorescence (SF)~\cite{Agarwal,Eberly,Macovei}.
Since the pioneering work of Dicke superradiance~\cite{Dicke}, many studies have confirmed 
superradiant effects in different samples. The first experimental observation of SF was 
realized by Skribanowitz \emph{et al}~\cite{Skribanowitz} in hydrogen fluoride gas.
Then, SF has been reported in a Cesium beam~\cite{Gibbs}, laser-cooled Rubidium 
atoms~\cite{coolRb} and Rubidium atomic vapor~\cite{Scully}. Very recently, giant superfluorescent 
bursts were observed in a dense semiconductor plasma~\cite{GiantSF}. Several further remarkable 
schemes for achieving ultrashort SF pulses have been demonstrated. In particular, superradiant 
amplification has been demonstrated in an underdense plasma~\cite{plasma}. The superradiance in 
a free-electron laser was observed too~\cite{FEL}. Furthermore, the work on SF indicates an 
important potential application for producing coherent, ultrashort and intense laser pulses, 
particularly, in the high frequency regions [ultraviolet (UV), extreme ultraviolet (EUV), and x-ray] 
by a suitable choice of the system and pump wavelength. The UV SF emission was reported in 
nanostructures at room temperature~\cite{UV}, whereas SF in the visible region has been characterized 
by pumping in the EUV region the helium gas~\cite{helium}. 

Such constructive interferences can also be observed when multilevel atoms are considered.
For instance, spontaneous emission can lead to so-called decay-induced coherence~\cite{Agarwal,Macovei}
between atomic dipole transitions under the stringent conditions on the level scheme in single atoms.
These conditions are not easily met for many schemes discussed in the literature.
Particularly, the well-known $\Lambda$- and V-type  three-level schemes with near-degenerate
nonorthogonal transition dipole moments can not be found in real atomic system.
Nevertheless, these spontaneous emission interferences do exist in some selected realistic 
level schemes, as for instances, four-level atom in the $J=1/2 \leftrightarrow J=1/2$ configuration 
in Mercury ions~\cite{Kiffner} or an atomic four-level system in the $N$-configuration which gives 
rise to electromagnetically induced absorption (EIA)~\cite{Taichenachev}.
Notably, the presence of spontaneous coherence was reported in artificial quantum systems 
like quantum wells~\cite{Faist,Schmidt} and quantum dots~\cite{Dutt}. Moreover, different approaches 
have been performed to induce spontaneous interference externally based on incoherent driving fields, 
coherent driving fields or breaking of the symmetry of the mode structure of the surrounding vacuum
~(see~\cite{Macovei} and references therein).
\begin{figure}[b]
\includegraphics[width=0.44\textwidth]{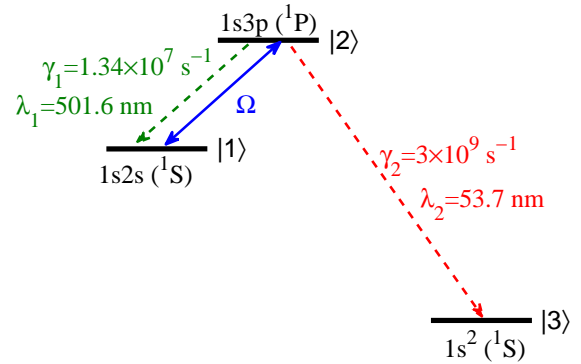}
\caption{\label{fig-1}(Color online)
The energy levels of a helium atom. 
The atomic transition $|2\rangle \to |1\rangle$
is coupled by a strong coherent laser field with Rabi frequency $\Omega$. 
$\gamma_{1}~(\lambda_1)$ and $\gamma_{2}~(\lambda_2)$ 
are the single-atom spontaneous decay rates (transition wavelength)
on transitions $|2\rangle \to |1\rangle$ and $|2\rangle \to |3\rangle$, respectively.}
\end{figure}

%
Here, we propose to realize EUV superfluorescence in an ensemble of helium atoms
coupled with a coherent laser field in the visible region. For this purpose, we employ a 
dense ensemble of $\Lambda-$type helium atoms where the emitters interact with each 
other via the common radiation field, see Fig.~\ref{fig-1}. In contrast to the work 
in Ref.~\cite{helium}, we use a coherent laser field to drive the visible transition 
$|1\rangle \leftrightarrow |2\rangle$, and investigate the cooperative spontaneous 
radiation on the EUV transition $|2\rangle \to |3\rangle$. Due to the constructive 
interatomic and intraatomic interferences, the superfluorescence in the EUV region takes 
place and can be controlled by the density of the gas and the Rabi frequency of the 
coherent driving field. In particular, due to laser induced quantum interference effects 
in the decay channels one can obtain two short SF pulses at the same frequency separated 
by a time delay. This could be useful in pump-probe experiments.

Our system is characterized by $N$ identical nonoverlapping $\Lambda$-type three-level helium atoms, 
consisting of $1s2s(^{1}S)$ state $|1\rangle$, $1s3p(^{1}P)$ state $|2\rangle$ and $1s^{2}(^{1}S)$ 
state $|3\rangle$, as shown in Fig.~\ref{fig-1}. The atomic transition between states $|1\rangle$ 
and $|2\rangle$ is resonantly coupled by a strong laser field with Rabi frequency $\Omega$,
and both transitions $|2\rangle \leftrightarrow |1\rangle$ and $|2\rangle \leftrightarrow |3\rangle$ 
are coupled via vacuum modes with orthogonal dipole moments $d_{21}$ and $d_{23}$, respectively. 
In the usual mean-field, Born-Markov and rotating-wave approximations, the interaction of the atomic 
system coupled by a resonant coherent laser field is described by the master equation:
\begin{eqnarray}
\dot{\rho}(t) =&-& i\Omega\sum_{j=1}^{N} [ S_{12}^{(j)},\rho ]
-\sum_{j,l=1}^{N}\{ \gamma_{jl}^{(1)}[S_{21}^{(j)},S_{12}^{(l)}\rho] \nonumber \\
&+& \gamma_{jl}^{(2)}[S_{23}^{(j)},S_{32}^{(l)}\rho]\} + \text{H.c.}.\label{meq}
\end{eqnarray}
Here $S^{(j)}_{\alpha\beta}=|\alpha\rangle_{jj}\langle\beta|$ with $\{\alpha,\beta\} \in\{1,2,3\}$ 
describe the population of the state $|\alpha\rangle$ in the $j$-th atom if $\alpha=\beta$ 
or the transition operator from $|\beta\rangle$ to $|\alpha\rangle$ if $\alpha\neq\beta$. 
These operators obey the commutation relations $[S^{(j)}_{\alpha\beta},S^{(l)}_{\beta'\alpha'}]=\delta_{jl}\left(\delta_{\beta\beta'}S^{(j)}_{\alpha\alpha'}- \delta_{\alpha\alpha'}S^{(j)}_{\beta'\beta} \right)$. Further, 
$\gamma_{jl}^{(i)}\equiv\gamma_{i}[\aleph_{jl}^{(i)} + i\Omega_{jl}^{(i)}]~(i\in\{1,2\})$
are the collective parameters \cite{Agarwal,Eberly,Macovei}, with $\aleph_{jl}^{(i)}$ and $\Omega_{jl}^{(i)}$
describing the mutual interactions among atomic pairs given by the expressions:
$\aleph_{jl}^{(i)}=\sin{(\omega_{i} r_{jl}/c})/(\omega_{i} r_{jl}/c)$ and
$\Omega_{jl}^{(i)}=-\cos{(\omega_{i} r_{jl}/c)}/(\omega_{i} r_{jl}/c)$,
where we have averaged over all dipole orientations. 
$\omega_{1}~(\omega_2)$ is the transition frequency on $|2\rangle\to|1\rangle ~(|2\rangle\to|3\rangle)$ 
transition. $r_{jl}=|\vec{r}_{j}-\vec{r}_{l}|$ is the interval between the $j$th atom and the $l$th atom.
$\gamma_{1}\approx 1.34\times 10^{7}~\rm s^{-1}$ and $\gamma_2\approx 3.0\times 10^{9}~\rm s^{-1}$ 
are the single-atom spontaneous decay rates from $|2\rangle \to |1\rangle$ and $|2\rangle \to |3\rangle$, 
respectively.

In the following, we investigate the collective spontaneous emission on the EUV 
transition $|2\rangle \to |3\rangle$ in an ensemble of helium atoms with a high 
density $n=10^{16}~\rm cm^{-3}$ with respect to the coherent laser field 
exciting the atoms from states $|1\rangle$ to states $|2\rangle$. We assume the 
atoms being initially prepared in the metastable states $|1\rangle$. The helium 
gas density is chosen as $n \sim \lambda_{2}^{-3}$ such that both transitions 
involved $|2\rangle \to |1\rangle$ and $|2\rangle \to |3\rangle$ are collective. 
The dynamics of the cooperative decay from $|2\rangle$ to $|3\rangle$ is 
investigated via numerical integration of the master equation~(\ref{meq}).
The equations for the population on the ground state $\langle S_{33}\rangle/N=
\sum^{N}_{j=1}\langle S^{(j)}_{33}\rangle/N$ and the intensity of SF emission 
$\langle S_{23}S_{32}\rangle/N^2=\sum^{N}_{j,l=1}\langle S^{(j)}_{23}S^{(l)}_{32}\rangle/N^2$ 
are governed by the number of collectively interacting helium atoms $N$,
the geometrical factor $\mu_{1}= \frac{3}{8\pi}(\frac{\lambda_{1}^{2}}{S})$, 
$\mu_{2}= \frac{3}{8\pi}(\frac{\lambda_{2}^{2}}{S})$ 
(when the ensemble of helium atoms has a form of a circular cylinder with length $L$ 
and cross sectional area $S$~\cite{Eberly}), decay rates $\gamma_{1,2}$ and Rabi frequency $\Omega$. 
We choose the sample parameters as $L=1~\rm mm$, $S=1~\rm mm^{2}$,
and the effective number of cooperating atoms $\mu_2 N\approx 3.4\times 10^{3}$.

\begin{figure}[t]
\includegraphics[width=0.22\textwidth]{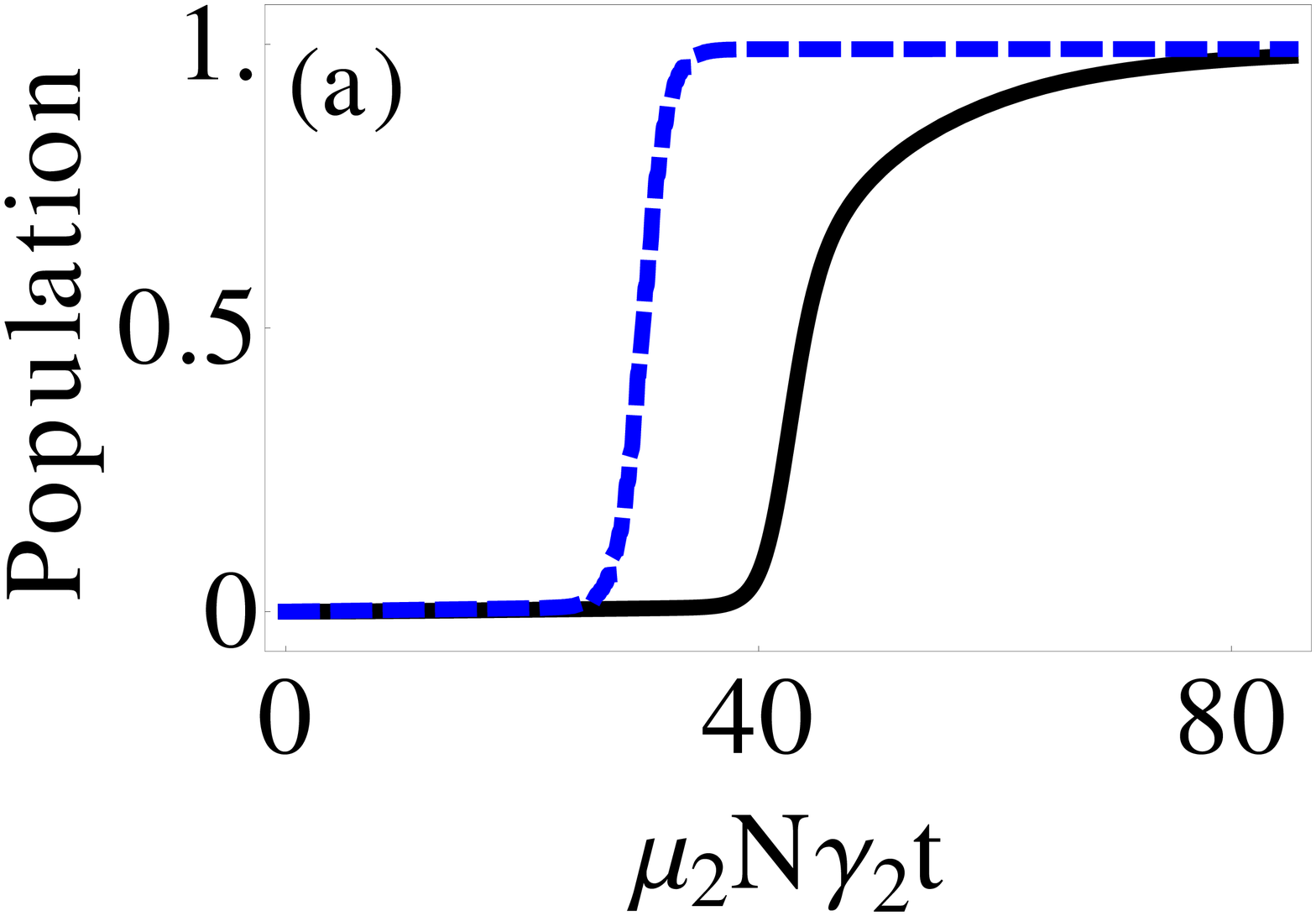}
\hspace{2.0mm}
\includegraphics[width=0.2205\textwidth]{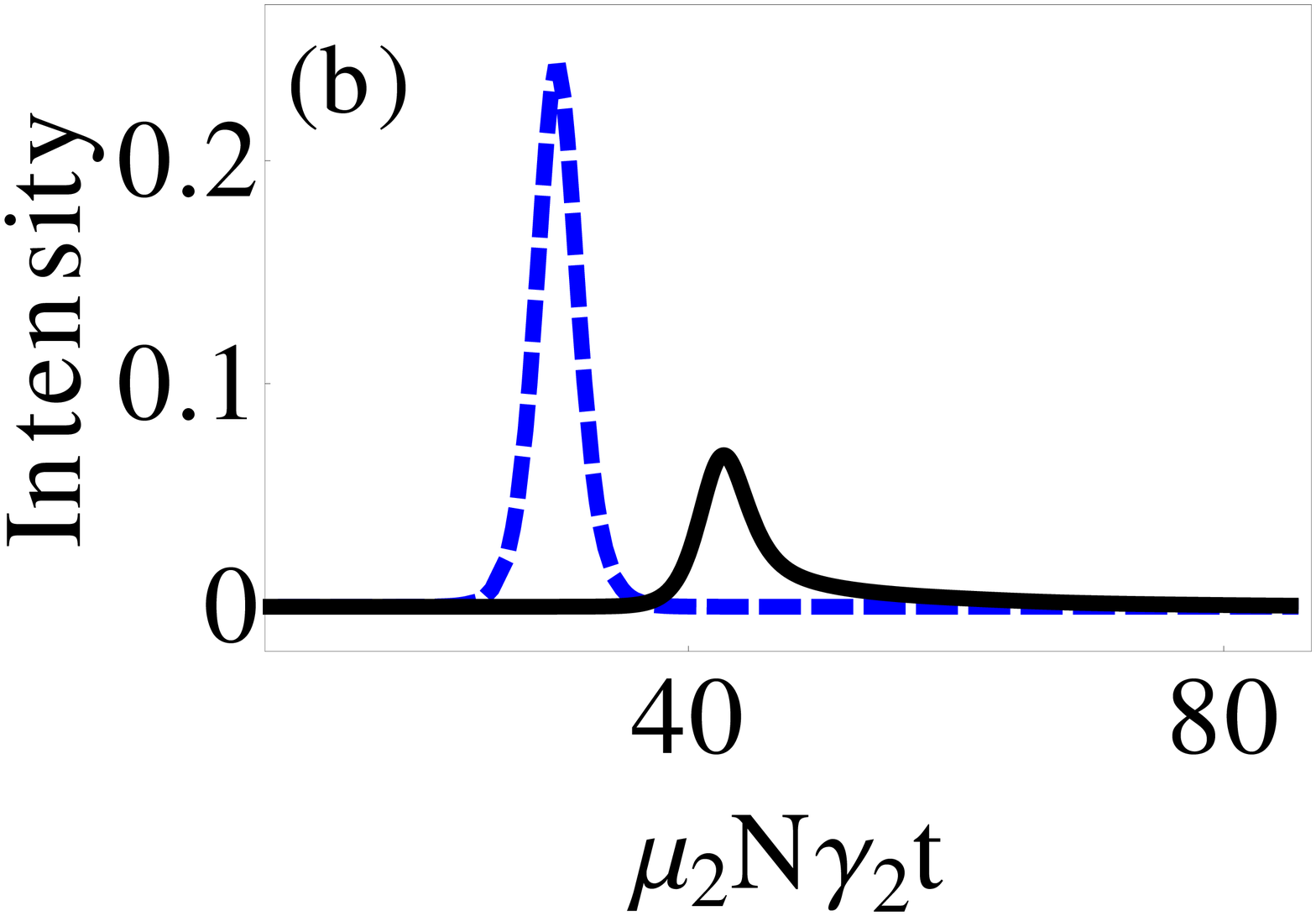}
\caption{\label{fig-2}(Color online) (a) The population of the ground state $|3\rangle$ 
and (b) the superfluorescence intensity (in units of $\mu_{2}\gamma_{2}N^{2}$) on the 
transition $|2\rangle \rightarrow |3\rangle$
as a function of the scaled time $\mu_{2}N\gamma_{2} t$ for different Rabi frequencies:
$\Omega\ll \mu_2 N\gamma_2$ (black solid curves) and $\Omega\gg \mu_2 N\gamma_2$ (blue dashed curves).
Here, $\mu_1/\mu_2\approx 87.3$ and $\gamma_1/\gamma_2\approx 1/223$.}
\end{figure}

First, we consider the case when the Rabi frequency $\Omega$ is not commensurate with 
the collective decay rates, i.e. $\Omega\ll \mu_{2} N \gamma_{2}$ or 
$\Omega\gg \mu_{2} N \gamma_{2}$. The evolution of population in the ground states 
$|3\rangle$ as well as the intensity of the emitted SF pulses is presented in Fig.~\ref{fig-2}. 
In both cases, the atoms in the metastable states $|1\rangle$ can be excited to the upper 
level state $|2\rangle$ by the coherent driving field. When a sufficient number of atoms 
is accumulated in state $|2\rangle$ those atoms will cooperatively decay to the ground states 
$|3\rangle$ in a short time and a SF burst takes place. However, there are some distinct 
differences for different strengths of the coherent driving field. In the case of a smaller 
Rabi frequency ($\Omega\ll \mu_{2} N \gamma_{2}$), the driving field is so weak that it 
needs more time to excite the atoms from the state $|1\rangle$ to the upper state 
$|2\rangle$ and thus the population in the ground state increases slowly [Black solid curve 
in Fig.~\ref{fig-2}(a)] with a weak SF burst [Black solid curve in Fig.~\ref{fig-2}(b)].
When $\Omega \gg \mu_{2} N \gamma_{2}$ [the blue dashed curve in Fig.~\ref{fig-2}] 
two pulses with higher intensities are emitted at different frequencies, i.e. 
$\omega_{23} \pm \Omega$, and their common intensity is shown in Fig.~\ref{fig-2}(b) 
with a blue dashed line.

\begin{figure}
\includegraphics[width=0.22\textwidth]{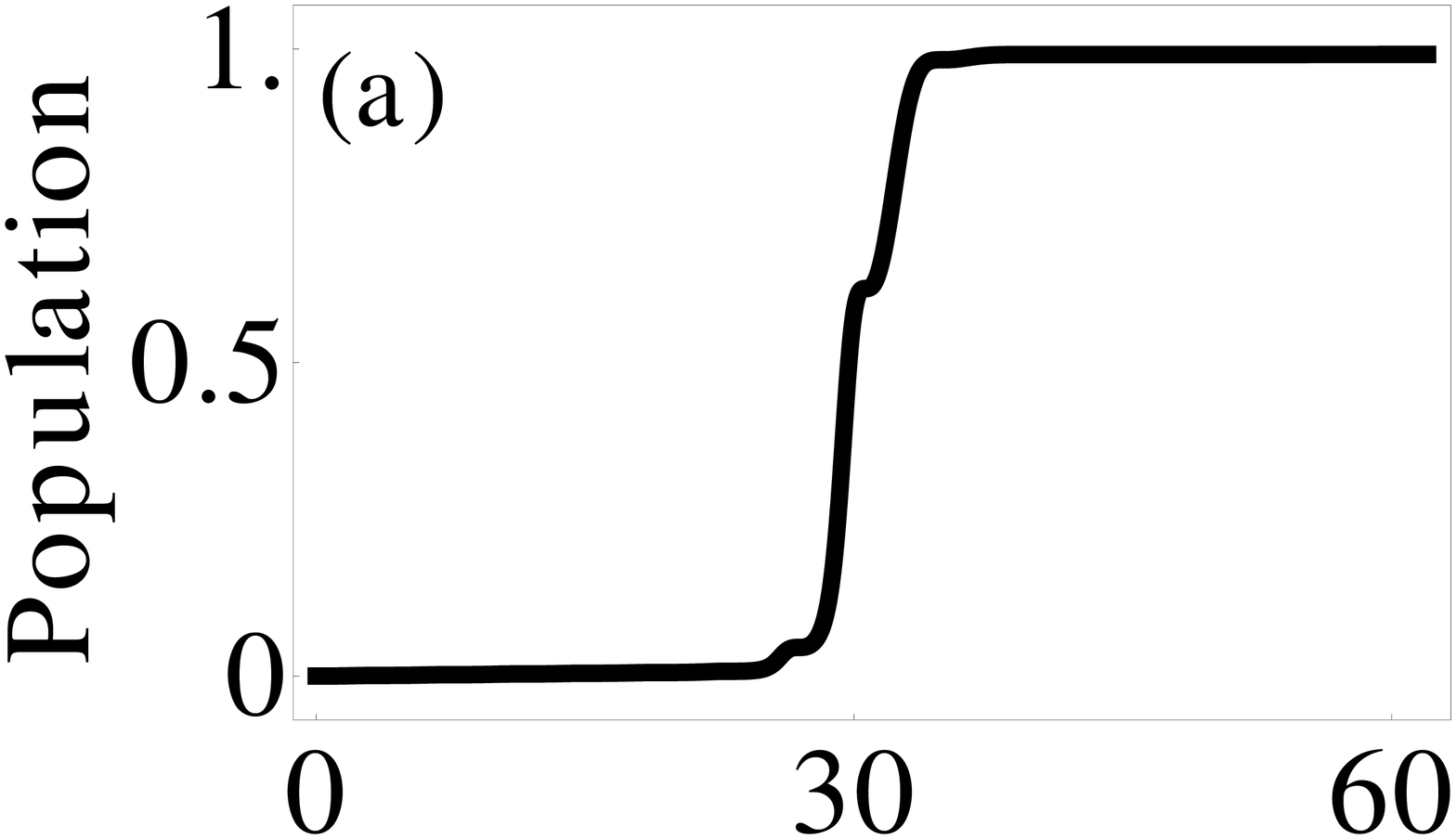}
\hspace{2.mm}
\includegraphics[width=0.2205\textwidth]{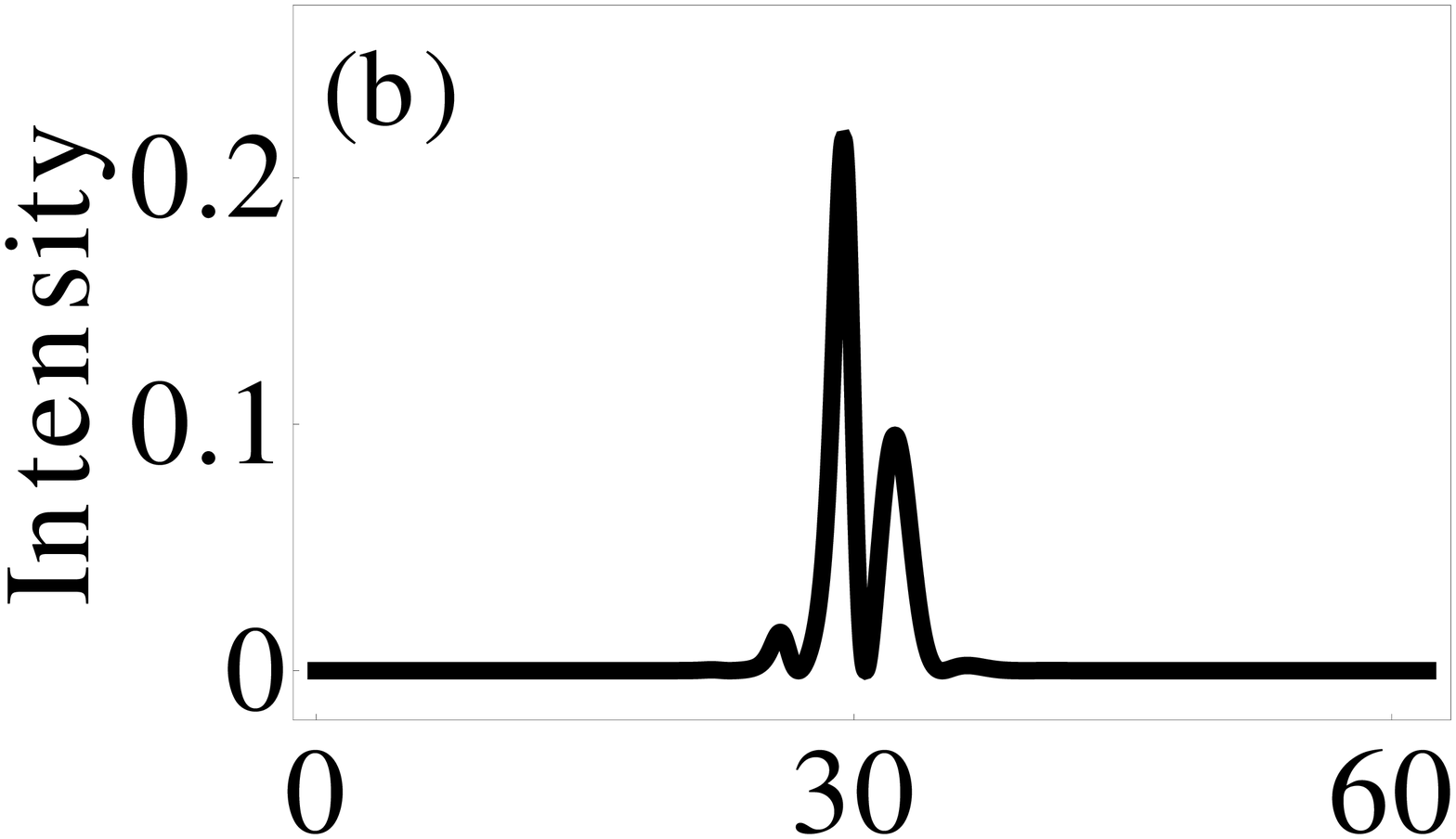}

\vspace{-4.mm}

\includegraphics[width=0.22\textwidth]{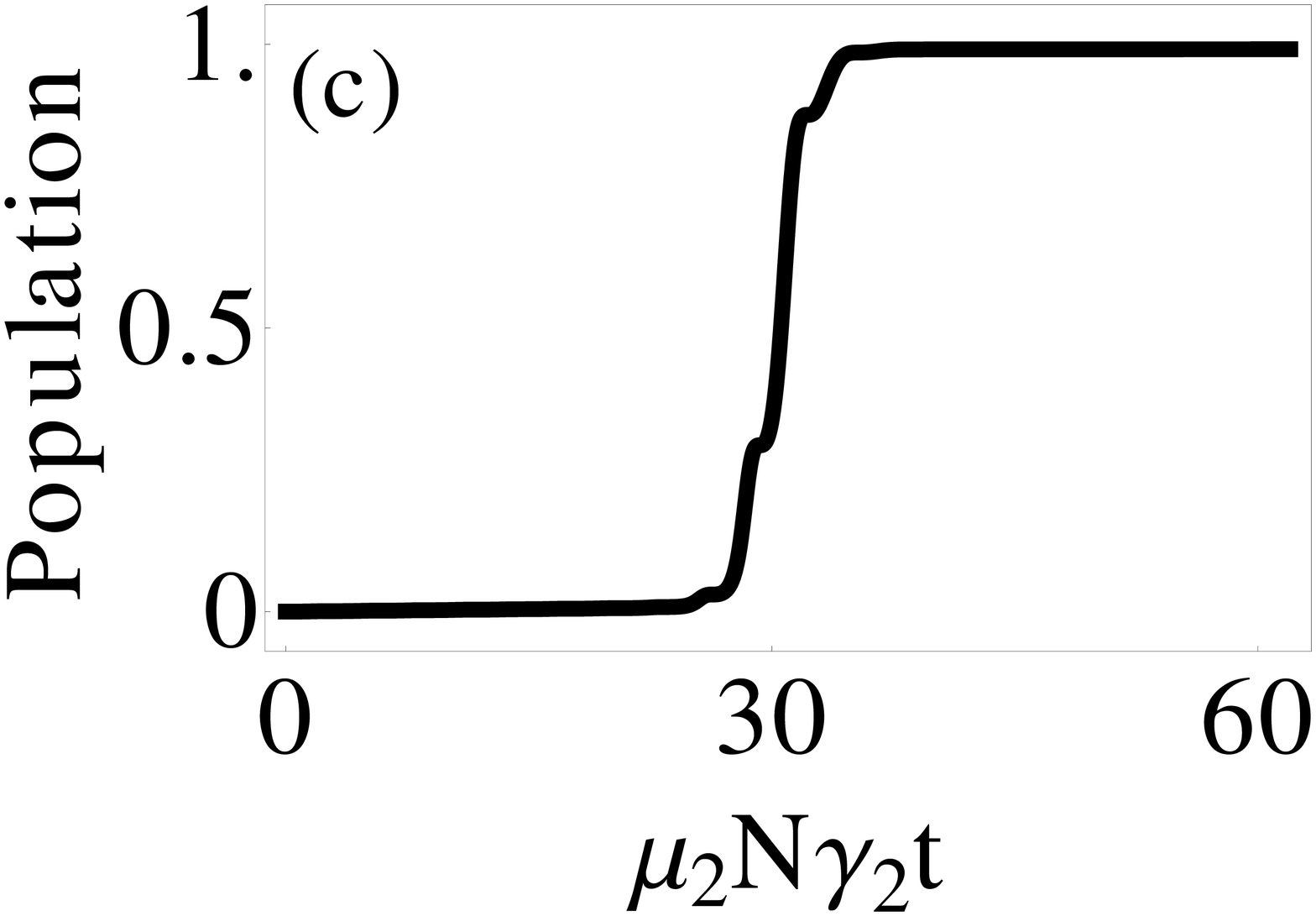}
\hspace{2.mm}
\includegraphics[width=0.2205\textwidth]{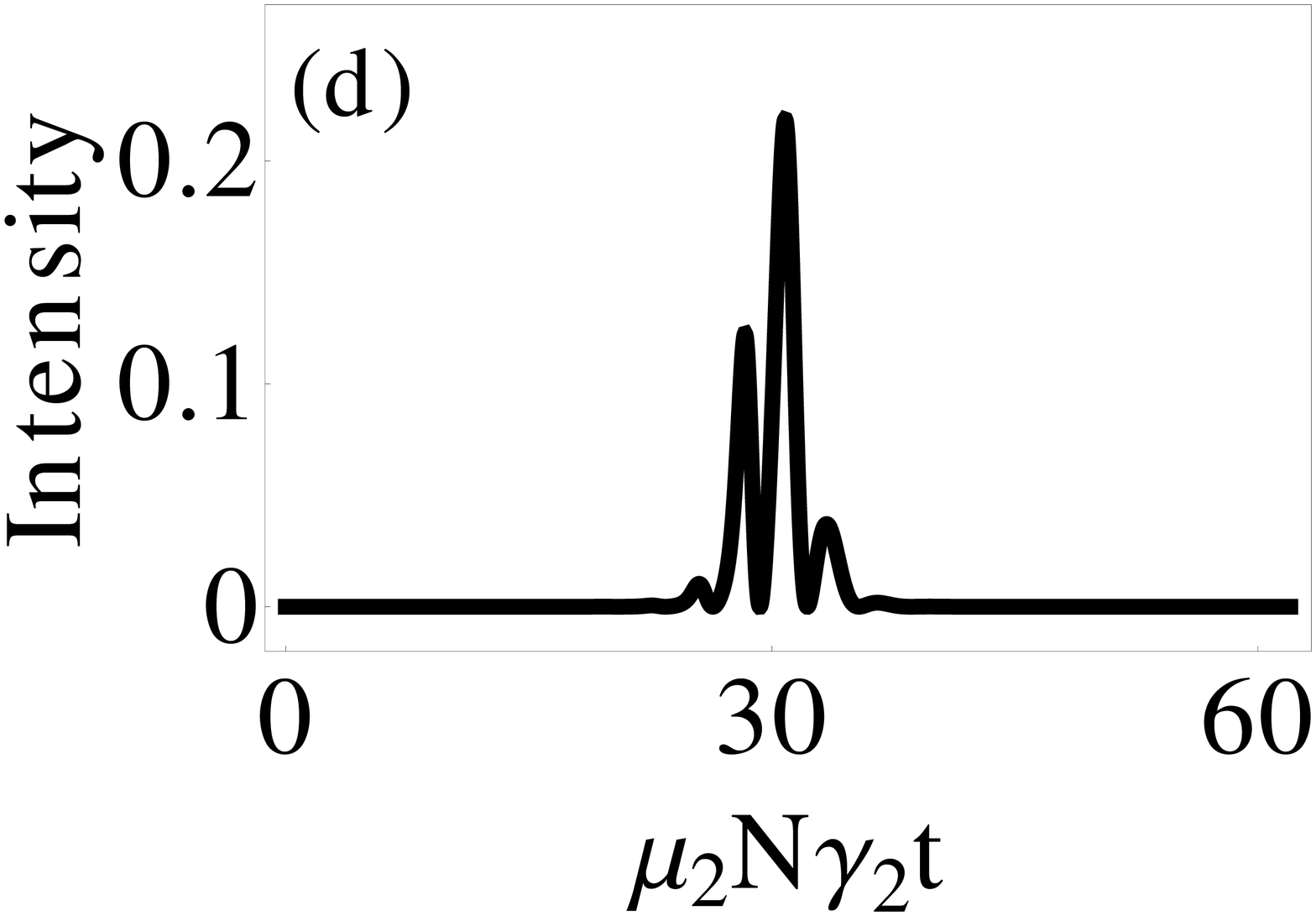}
\caption{\label{fig-3} (a), (c) The population in the ground state $|3\rangle$ 
and (b), (d) the superfluorescence intensity (in units of $\mu_{2}\gamma_{2}N^{2}$)
on the transition $|2\rangle \rightarrow |3\rangle$ as a function of the scaled time 
$\mu_{2}N\gamma_{2} t$. (a), (b) $\Omega=0.84\mu_2 N\gamma_2$ whereas (c), (d) 
$\Omega=1.09\mu_2 N \gamma_2$. All other parameters are the same as in Fig.~\ref{fig-2}.}
\end{figure}

%
Now, we turn to the more interesting case $\Omega\sim\mu_{2}N\gamma_2$.
Figure~(\ref{fig-3}) depicts the evolution of populations in the ground states $|3\rangle$ 
and the intensity of SF emission for the Rabi frequency $\Omega=0.84\mu_2 N\gamma_2 
\approx 8.57 \times 10^{12}$Hz and $\Omega=1.09\mu_2 N \gamma_2 \approx 11.12\times 10^{12}$Hz. 
It is shown that, after the excitation from the metastable state $|1\rangle$ to the upper 
states $|2\rangle$ by the driving field $\Omega$, the atoms eventually decay cooperatively to 
the ground states. The population in ground state $|3\rangle$ increases to unity in a short 
time with some small modulations due to the SF bursts. There are evidently two short pulses of 
the same frequency with a time delay emitted during the cooperative spontaneous emission processes, 
but the SF pulses display quite different features for different Rabi frequencies. For 
$\Omega=0.84\mu_2 N\gamma_2$, the leading pulse is much stronger than the delayed one, while the 
feature is opposite when $\Omega=1.09\mu_2 N\gamma_2$. 
Thus, there is a direct link between the emitted SF pulses and the Rabi frequency $\Omega$ that,
in particular, can be used to control the generation of the SF pulses.
The two pulses in Fig.~\ref{fig-3}(b) can be used for pump-probe experiments in EUV domain.

%
The phenomena above can be physically understood in the dressed-state picture.
Due to the strong coupling via the driving field, the upper state $|2\rangle $ 
splits into two states $|+\rangle$ and $|-\rangle$ separated by $2\Omega$. The 
two states will both decay to the ground state $|3\rangle$ with the rates 
proportional to $\gamma_{2}^{(+)}\approx\gamma_{2}^{(-)} \equiv \gamma_{2}$, 
respectively. When the Rabi frequency $\Omega$ is large,  the atoms in states 
$|+\rangle$ and $|-\rangle$ would independently decay to the ground state 
$|3\rangle$ by emitting two pulses with different frequencies $\omega_{23} \pm \Omega$.
However, for the smaller Rabi frequency $\Omega \sim \mu_{2}N\gamma_{2}$, 
the two states $|+\rangle$ and $|-\rangle$ are so close that the decay amplitudes to 
the ground state interfere with each other, similar to the case of a $V$-type 
three-level scheme with near-degenerate nonorthogonal transition dipole moments
~\cite{Agarwal,Macovei}. These decay-induced coherences may give rise to 
destructive quantum interference between the two decaying paths leading to the 
splitting of the SF pulse with a time delay proportional to $\Omega^{-1}$ shown 
in Fig.~\ref{fig-3}(b,d). The ionization probability of the 1s3p excited state 
via one-photon processes can be estimated using the cross-section from \cite{ion}, 
i.e. $\sigma_{i}\approx 10^{-17}{\rm cm}^{2}$ at $\lambda_{1}\approx 500{\rm nm}$. 
At laser intensities $I_{L} \sim 10^{9}-10^{10}{\rm W/cm^{2}}$ and interaction times 
$\tau \sim 2\pi/\Omega$ the ionization probability 
$P_{i}=\sigma_{i}I_{L}\tau/(\hbar \omega_{1})$ is of the order of $10^{-2}-10^{-1}$ 
and, thus, negligible small. 

Finally, we demonstrated the generation of SF in the EUV region in an ensemble of $\Lambda$-type 
three-level helium atoms. However, the technique could be easily extended to other atomic 
or atomic-like systems with the $\Lambda$-type three-level structures, as for instance, 
He-like Carbon $\rm C^{4+}$, that could be used to produce X-ray SF if high densities are 
available. 

In summary, we have shown that the cooperative spontaneous emission arising at high 
density of a helium atomic ensemble can be employed to generate SF pulses in the EUV 
region. In particular, the collective effects for a helium ensemble coupled 
simultaneously with a proper laser field are shown to induce quantum interference in EUV SF leading to a 
preponderate two-pulse emission with a potential application for pump-probe experiments.

We benefited from useful discussions with Karen Z. Hatsagortsyan.


\newpage

\end{document}